\documentclass[12pt]{iopart}
\usepackage{iopams}  
\usepackage{graphicx}
\usepackage{bm,bbm}
\usepackage{xcolor}

\newcommand{\be}{\begin{equation}}
\newcommand{\ee}{\end{equation}}

\newcommand{\ba}[1]{\left(\begin{array}{#1}}
\newcommand{\ea}{\end{array}\right)}
\begin{document}
\title[Heat exchange and fluctuation in Gaussian thermal states]{Heat exchange and fluctuation in Gaussian thermal states in the quantum realm}

\author{A. R. Usha Devi$^{1,2}$, Sudha$^{3,2}$, 
A. K. Rajagopal$^2$ and A. M. Jayannavar$^4$}

\address{$^1$ Department of Physics, Jnanabharathi, Bangalore University, Bangalore-560056, India}
\address{$^2$ Inspire Institute Inc., Alexandria, Virginia, 22303, USA.}
\address{$^3$ Department of Physics, Kuvempu University, 
	Shankaraghatta-577 451, India}
\address{$^4$ Institute of Physics, 
Bhubaneshwar, India}
\ead{arutth@rediffmail.com}

\begin{abstract} 
The celebrated exchange fluctuation theorem -- proposed by Jarzynski and W{\'o}zcik, (Phys. Rev. Lett. {\bf 92}, 230602 (2004)) for heat exchange between two  systems in thermal equilibrium at different temperatures -- is explored here for quantum Gaussian states in thermal equilibrium.  
We employ Wigner distribution function formalism for quantum states, which exhibits close resemblance with the classcial phase-space trajectory description, to arrive at this theorem. For two Gaussian states in thermal equilibrium at two different temperatures  kept in contact with each other for a fixed duration of time we show that the quantum  Jarzyinski-W{\'o}jcik theorem agrees with the corresponding classical result in the limit $\hbar\rightarrow 0$. 
\end{abstract}

\maketitle
\section{Introduction} 
Fluctuation theorems~\cite{j97,j97a,c98,c99,jw}
are of fundamental significance in non-equilibrium statistical physics. They correspond to a collection of exact relations, which remain valid even when the system is driven far away from equilibrium.  Various exchange-fluctuation theorems (XFT)   involving thermodynamic quantities like work, heat, entropy have been proposed during the last two decades~\cite{jw, phqe, Dhar07, xft09,rmp9,ph10,rmp11,
	pre12,prjp12,prx13,prjp13,epjb14,ijp15,ppnature15, pre15,pre17,Lutz18,fqrx18,pre18,rmp20,prl20}. They have offered significant insights in understanding  thermodynamical processes -- especially the emergence of irreversibility from reversible dynamics and the directionality of heat flow implied by second law of thermodynamics. Some of these relations are applicable for systems in non-equilibrium steady state~\cite{epjb14}, while  others hold in the transient regime. Fluctuation relations with underlying Hamiltonian dynamics~\cite{phqe,prx13} as well as stochastic dynamics~\cite{rmp11,prjp12,prjp13,ijp15} have also been proposed. There are ongoing efforts to generalize and broaden the applicability of  XFT in the quantum scenario ~\cite{rmp9,rmp11, pre15,fqrx18,rmp20,prl20}.
	
	The fluctuation-exchange relations can be considered as generalizations of second law of thermodynamics for small systems and they connect the probabilities of appearance of physical quantitites such as work, heat, number of particles, in an experimental set up, to those obtainable in a time-reversed set up. For instance,  the Jarzynski-W{\'o}zcik fluctuation theorem~(XFT)~\cite{jw} given by, 	
	\begin{eqnarray}
	\label{eq1}
	\ln\left[\frac{p_\tau(+{\cal Q})}{p_\tau(-{\cal Q})}\right]&=&\bigtriangleup\beta \, {\cal Q}, \ \ \ \bigtriangleup\beta=\frac{1}{k\,T_{B}}-\frac{1}{k\,T_{A}}
	\end{eqnarray}    
	quantifies the ratio of probability $p_\tau(+{\cal Q})$ of heat exchange during interaction of  $A$ and $B$ for a fixed time duration  $\tau$, to its time-reversed counterpart $p_\tau(-{\cal Q})$. Here $k$ denotes Boltzmann constant and ${\cal Q}$ denotes the amount of heat exchanged.   
	          
The Jarzynski-W{\'o}zcik theorem~\cite{jw} is one among the important XFTs and has drawn much attention. 
Microreversibility and  strict directionality of thermodynamical heat flow form the foundational features of this XFT relation. 
 Generalizations of the theorem  to include processes involving a system coupled to reservoirs~\cite{pre17}, a chain of interacting particles connecting two heat baths~\cite{epjb14}, correlated thermal quantum systems~\cite{pre15,prl20} and the like have been carried out. 

In the classical scenario Jarzynski and W{\'o}zcik~\cite{jw} had employed  phase-space description, for the forward and reverse dynamical evolution between statistical systems $A$ and $B$ in thermal equilibriums at  temperatures $T_A$ and  $T_B$ respectively, so that one gains physical intuition underlying the relation between heat exchange and fluctuations. In the quantum regime, they considered systems with discrete energy levels to arrive at the relation (\ref{eq1}).  

In the present work, we retain the flavour of the phase-space approach in the quantum scenario,  by confining ourselves to continuous variable Gaussian thermal states. Following similar lines as that of the original work~\cite{jw} we arrive at the heat exchange-fluctuation theorem in the quantum realm for two Gaussian states in thermal equilibrium at temperatures $T_A$, $T_B$ kept in contact with each other.  What comes to our aid here is the fact that the Wigner distribution function, characterizing  Gaussian states is non-negative~\cite{hil} and hence, it serves as a {\em legitimate} quantum counterpart of phase-space probability distribution. This enables us to carry out explicit evaluations and arrive at the Jarzynski-W{\'o}zcik XFT in this case.  

We have organized our paper as follows: In Section~2 a brief outline of Jarzynski-W{\'o}zcik derivation of heat transfer and fluctuation relation (\ref{eq1}) is presented. Necessary mathematical preliminaries on infinite dimensional continuous variable  Gaussian systems, their characterization in terms of the variance matrix and  obtaining the Wigner distribution function in terms of the variance matrix are given in Section~3. In Section~4, we derive the  Jarzynski-W{\'o}zcik  heat exchange-fluctuation theorem for  Gaussian systems $A$ and $B$, in thermal equilibrium at temperatures $T_A$, $T_B$ respectively using the Wigner distribution function approach. Time reversal symmetric canonical transformations of phase-space observables  is employed to identify explicit forms of forward and backward heat probability distributions  using the Wigner distribution function associated with Gaussian thermal states (or equivalently,  quantum harmonic oscillator system in thermal equilibrium).  Discussions on heat exchange statistics of quantum and classical harmonic oscillators in thermal equilibrium,  physical status of the Wigner-Weyl phase-space trjectory framework and interpretation of the quantum-to-classical reduction in the limit $\hbar\rightarrow 0$, possible connection of Jarzynski-W{\'o}zcik XFT with energy equipartition theorem in the quantum scenario are presented in Section~5.    

\section{Classical phase-space description for Jarzynski-W{\'o}zcik heat exchange-fluctuation relation}

Jarzynski and W{\'o}zcik considered two systems, phase-space evolution of which is governed by Hamiltonians  $H_A(\xi_A)$ and $H_B(\xi_B)$;  $\xi_A$, $\xi_B$ denoting phase-space variables (e.g., positions and momenta) of systems $A$ and $B$ respectively. The systems are kept in contact  with each other for a time duration $\tau$ via an  interaction characterized by $H_{\rm int}(\xi_A,\xi_B)$, which is switched `on' at time $t=0$, and turned `off' at  $t=\tau$. The phase-space trajectory of the two systems is denoted collectively by   $\xi^t=\left(\xi^t_A,\xi^t_B\right)$.  
Both the systems are initially in thermal equilibrium, at temperatures $T_A$, $T_B$ respectively, and their phase-space probability distributions at time $t=0$ is given by 
 \begin{eqnarray}
 \label{prob}
 p(\xi^0)=\frac{e^{-H_A(\xi^0_A)/k\,T_A}\, e^{-H_B(\xi^0_B)/k\,T_B}}{Z_A\, Z_B} 
 \end{eqnarray}
 where $Z_A,\ Z_B$ denote partition functions. Phase-space dynamics of the systems is assumed to be time-reversal symmetric i.e., 
 \begin{eqnarray}
 H_A(\xi_A)\longrightarrow H_A(\xi^*_A) = H_A(\xi_A) \nonumber \\
 H_A(\xi_B)\longrightarrow H_B(\xi^*_B) = H_B(\xi_B)  \\
 H_{\rm int}(\xi)\longrightarrow  H_{\rm int}(\xi^*)=H_{\rm int}(\xi) \nonumber
 \end{eqnarray} 
 where time reversal operation is denoted by the superscript symbol $(\, ^*\, )$.  In other words, for every legitimate forward trajectory  $\xi^0$ to $\xi^\tau$, there exists a time-reversed  trajectory $\bar{\xi}^0=\xi^{\tau*}$ to $\bar{\xi}^\tau=\xi^{0*}$. Their likelihood ratio is given by (see (\ref{prob})) 
 \begin{eqnarray}
 \label{jwcl}
 \frac{ p(\xi^0)}{ p(\bar{\xi}^0)}&=&e^{\left(H_A(\bar{\xi}^{0}_A)-H_A(\xi^0_A)\right)/k\,T_A}\, e^{\left(H_B(\bar{\xi}^{0}_B)-H_B(\xi^0_B)\right)/k\,T_B} \nonumber \\
    &=& e^{\Delta\,E_A/k\,T_A}\, e^{\Delta\,E_B/k\,T_B} 
 \end{eqnarray} 
 where $\Delta\,E_A=H_A(\xi^{\tau*}_A)-H_A(\xi^{0}_A), \Delta\,E_B=H_B(\xi^{\tau*}_B)-H_B(\xi^{0}_B)$ denote change of internal energies of systems $A$ and $B$ respectively. Assuming that the  interaction term $H_{\rm int}$ is negligible, it is seen that   $H_A(\xi^{0}_A) + H_B(\xi^{0}_B)\approx H_A(\xi^{\tau*}_A) + H_B(\xi^{\tau*}_B)$
 or $\Delta\,E_A\approx -\Delta\,E_B$. Net energy change during the interaction represents the amount of heat transferred  i.e.,  ${\cal Q}=\Delta\,E_B\approx -\Delta\,E_A$. The heat transfer ${\cal Q}$ from $A$ to $B$  during forward process gets compensated by that in the reverse process from $B$ to $A$ and is expressed by      
 \begin{equation}
 \label{heatbf}
 {\cal Q}(\xi^0)=-{\cal Q}(\bar{\xi\,}^0).
 \end{equation}           
 Thus, it is seen that     
  \begin{eqnarray}
 \label{jwcl2}
 \frac{ p(\xi^0)}{ p(\bar{\xi}^0)}&=& e^{\Delta \beta\, {\cal Q}(\xi^0)}. 
 \end{eqnarray} 
From (\ref{heatbf}) and (\ref{jwcl2}) it follows that  
 \begin{eqnarray}
\label{Qdist}
p_\tau({\cal Q})&=&\int\, d\xi^0\, p(\xi^0)\, \delta(\xi^0-{\cal Q}) \nonumber \\ 
p_\tau(-{\cal Q})&=&\int\, d\bar{\xi}^0\, p(\bar{\xi}^0)\, \delta(\bar{\xi\,}^0+{\cal Q}) \nonumber \\  
     &=& e^{-\Delta\beta\, {\cal Q}}\, p_\tau({\cal Q})  \nonumber \\ 
     &\Longrightarrow&\  \ \ \ \ \ \frac{p_\tau({\cal Q})}{p_\tau(-{\cal Q})}= e^{\Delta\beta\, {\cal Q}}. 
\end{eqnarray}           
thus proving the Jarzynski-W{\'o}jcik heat exchange fluctuation theorem in the classical scenario. 
 
In the quantum realm Jarzynski and W{\'o}zcik considered two discrete level systems  prepared initially in thermal equilibrium at temperatures $T_A$, $T_B$ and measure their energies $E^A_i,\, E^B_i$; the systems are allowed to interact weakly for a time duration $\tau$ after interaction is turned off and  energies of both the systems $E^A_f$, $E^B_f$ measured. As the systems are allowed to interact weakly it is expected that the total energy of the system is conserved: $E^A_i+E^B_i\approx E^A_f+E^B_f$.  
Heat transfer is then interpreted as ${\cal Q}_{i\rightarrow f}= E^B_i-E^B_f\approx E^A_f-E^A_i$ resulting in the relation  
\begin{eqnarray}
\label{jwq}
\ln\left[\frac{p\left(\vert i\rangle \stackrel{\tau}{\longrightarrow}\vert f\rangle\right)}{p\left(\vert f\rangle\stackrel{-\tau}{\longrightarrow}\vert i\rangle\right)}\right]=\bigtriangleup\beta\, {\cal Q}_{i\rightarrow f}.
\end{eqnarray} 
    
Our interest here is to derive the relation (\ref{Qdist}) describing heat transfer processes in the forward and the time-reversed dynamics of quantum Gaussian system consisting of two subsystems $A$, $B$,  prepared initially   in thermal equilibrium at temperatures $T_A$, $T_B$ respectively. Wigner distribution function formalism~\cite{hil} is  employed in this approach. To this end,  we give necessary mathematical preliminaries on the quantum phase-space  description and symplectic evolution, Gaussian thermal states and  the associated  Wigner distribution function in Section~3.       

\section{Gaussian states --  description through variance matrix and Wigner function}  

Gaussian states, the most important among continuous variable states~\cite{nm,nm1}, find diverse applications in several fields including quantum stochastic processes and open system dynamics~\cite{adesso}. They naturally occur as the thermal equilibrium states of any physical system in the small oscillations limit~\cite{adesso}. Being fully characterized by its first and second moments Gaussian states are simpler to handle among the continuous variable states.   
 
Any arbitrary two-mode Gaussian state ${\hat\rho}_{AB}$ is characterized completely by its first and second order moments, written concisely in the form of a  $4\times 4$ covariance matrix ${\bf V}$ (referred to as variance matrix from now on), which is  defined in terms of its  elements as~\cite{nm,nm1} 
\be
\label{var1}
V_{ij}=\frac{1}{2}\, \left\langle \left\{\hat{\xi}_i,\hat{\xi}_j\right\}\right\rangle-\langle \hat{\xi}_i\rangle \langle \hat{\xi}_j\rangle, \ \ i,\,j=1,2,3,4.
\ee
Here $\hat{\xi}_i$ denotes a $4\times 1$ column  ${\bf {\hat{\xi}}}$ with positions and momenta (dimensionless) as its components:  
\be
\label{xi}
{\bf {\hat{\xi}}}=(\hat{q}_A,\,\hat{p}_A,\,\hat{q}_B,\,\hat{p}_B)^{\rm T}, 
\ee
where `T' stands for the transpose operation; we have denoted $\left\{\hat{\xi}_i,\hat{\xi}_j\right\}=\hat{\xi}_i\hat{\xi}_j+\hat{\xi}_j\hat{\xi}_i$  and we have denoted  $\langle \cdots \rangle=\mbox{Tr}\,\left({\hat {\rho}}\cdots \right)$ in (\ref{var1}).   
The canonical phase-space variables $\hat{q}_\alpha$, $\hat{p}_\alpha$, $\alpha,\ \beta=A,\,B$ satisfy the Bosonic commutation relations, 
\begin{eqnarray}
\label{com}
\left[\hat{q}_\alpha,\,\hat{q}_\beta \right]=0,\ & & \left[\hat{p}_\alpha,\,\hat{p}_\beta \right]=0, \ \ \mbox{and}\nonumber \\ 
\left[\hat{q}_\alpha,\,\hat{p}_\beta \right]=i\, \delta_{\alpha\beta}; \ \ 
\end{eqnarray}
where $\delta_{\alpha\beta}=1$ when $\alpha=\beta$ and zero when $\alpha\neq \beta$, is the Kronecker delta function. In terms of the components $\hat{\xi}_i$, $i=1,2,3,4,$ the canonical commutation relations (\ref{com}) assume the form  
\be
\label{com2}
[\hat{\xi}_i,\,\hat{\xi}_j]=i\, \Omega_{ij}, \ i,\,j=1,\,2
\ee
where $\Omega_{ij}$ denote elements of the $4\times 4$  matrix 
\begin{eqnarray}
\label{om} 
{\bf \Omega}= \left(\begin{array}{cccc}  
0 & 1 & 0 & 0 \\ 
-1 & 0 & 0 & 0 \\ 
0 & 0 & 0 & 1 \\ 
0 & 0 & -1 & 0 
\end{array}\right).   
\end{eqnarray} 
The commutation relations (see  (\ref{com}), (\ref{com2})) remain invariant under a {\emph{symplectic}} transformation~\cite{nm,nm1}:  
\begin{equation}
\label{sp4r}
{\bf S}\,{\bf \Omega}\,{\bf S}^T={\bf \Omega}. 
\end{equation}
 The set of all $4\times 4$ real matrices $\bf S$ satisfying the property (\ref{sp4r}) constitutes the symplectic group of real linear canonical transformations Sp(4,R)~\cite{nm, nm1}.

The variance matrix ${\bf V}$ of a two-mode quantum system  given explicitly by 
\begin{eqnarray}
\label{cov} 
{\bf V}= \left(\begin{array}{cccc}  
\langle q^2_A\rangle & \frac{1}{2}\langle \{q_A,\,p_A\}\rangle & \langle q_A\,q_B\rangle & \langle q_A\,p_B\rangle \\
\frac{1}{2}\langle \{q_A,\,p_A\}\rangle & \langle p^2_A\rangle &  \langle p_A\,q_B\rangle & \langle p_A\,p_B\rangle \\
\langle q_A\,q_B\rangle & \langle q_B\,p_A\rangle &  \langle q^2_B\rangle & \frac{1}{2}\langle \{q_B,\,p_B\}\rangle \\
\langle q_A\,p_B\rangle & \langle p_A\,p_B\rangle &  \frac{1}{2}\langle \{q_B,\,p_B\}\rangle & \langle p_B^2\rangle \\
\end{array}\right)   
\end{eqnarray} 
is a real symmetric positive definite  matrix and it completely characterizes a two mode Gaussian state ${\hat\rho}_{AB}$.   Under the symplectic transformation $\hat{\xi}'={\bf S}\hat{\xi}$,  the variance matrix ${\bf V}$  undergoes a congruent transformation ${\bf V}'={\bf S}{\bf V}{\bf S}^T$, where ${\bf V}'$ is the variance matrix associated with the new canonical variables $\hat{q}\,'_\alpha$, $\hat{p}\,'_\beta$, $\alpha,\ \beta=A,\, B$.
 
From the fundamental theorem due to  Williamson~\cite{wil} it follows that the variance matrix $\bf V$ attains a canonical form under symplectic transformation 
${\bf S}_{\rm W}$ such that   
\be
{\bf V}_{\rm W}={\bf S}_{\rm W}{\bf V}\,{\bf S}_{\rm W}^T=\mbox{diag}\,(\nu_A,\,\nu_A;\, \nu_B,\,\nu_B)
\ee
and  ${\bf V}_{\rm W}$ is referred to as the {\emph {Williamson normal form}} of the variance matrix and $\nu_A,\, \nu_B$ are called the symplectic eigenvalues of the variance matrix. The real positive diagonal elements  $\nu_\alpha$, $\alpha=A,\, B$ of ${\bf V}_{\rm W}$ are the  positive square roots of the doubly degenerate eigenvalues of the matrix $-\left({\bf V \Omega}\right)^2$ as,   
\be
-{\bf S}_{\rm W} \left({\bf V \Omega}\right)^2{\bf S}_{\rm W}^{-1}=-\left({\bf V}_{\rm W}{\bf \Omega}\right)^2=\mbox{diag}\,(\nu_A^2,\,\nu_A^2,\, \nu_B^2,\,\nu_B^2).
\ee
Corresponding to the symplectic transformation ${\bf S}_{\rm W}$, there exists a unitary operator ${\bf{U}}({\bf S}_{\rm W})$ transforming the density matrix 
$\hat{\rho}_{AB}$ of a two mode Gaussian state as follows: 
\be
\label{G2mode}
\hat{\rho}_{AB}={\bf{U}}^\dagger({\bf S}_{\rm W})\,\left( \hat{\rho}_{\nu_A}\otimes \rho_{\nu_B}\right)\,  {\bf{U}}({\bf S}_{\rm W}) 
\ee
where the single mode density matrices  ${\hat{\rho}}_{\nu_A},\, {\hat{\rho}}_{\nu_B}$  are given by~\cite{adesso} 
\be
\label{single} 
\hat{\rho}_{\nu_\alpha}=
\frac{1}{\nu_\alpha+\frac{1}{2}}\sum_{n_\alpha=0}^\infty\, 
\left(\frac{\nu_\alpha-\frac{1}{2}}{\nu_\alpha+\frac{1}{2}}\right)^{n_\alpha}\vert n_\alpha \rangle\langle 
n_\alpha \vert, \ \ \alpha=A,B. 
\ee  
Here  $\vert\, n_A\, \rangle,\, \vert\, n_B\,\rangle$ are the eigenstates of the number operators 
$\hat{N}_A=\hat{a}^\dagger_A\, \hat{a}_A, \ \hat{N}_B=\hat{a}^\dagger_B\, \hat{a}_B $ of the modes $A,B$ and   $\hat{a}_\alpha, \hat{a}_\alpha^\dag$ are related to the dimensionless  canonical position and momentum observables as follows:   
\begin{eqnarray}
\label{ac}
\hat{a}_\alpha&=& \frac{{\hat q}_\alpha + i\, {\hat p}_\alpha}{\sqrt{2}},\ \ \ \hat{a}^\dagger_\alpha=\frac{{\hat q}_\alpha - i\, {\hat p}_\alpha}{\sqrt{2}}.   
\end{eqnarray}

Let us denote $\hat{Q}_\alpha=\left(\frac{\hbar}{m_\alpha\,\  \omega_\alpha}\right)^{1/2}\, \hat{q}_\alpha, \ \ \hat{P}_\alpha=  \left(m_\alpha\, \omega_\alpha\, \hbar\right)^{1/2}\, \hat{p}_\alpha$. Given the Hamiltonian of a harmonic oscillator of mass $m_\alpha$,  frequency $\omega_\alpha$, 
\begin{eqnarray}
\label{thh0}
\hat{H}_\alpha&=&\frac{\hat{P}^2_\alpha}{2m_\alpha}+\frac{1}{2}\, m_\alpha\, \omega^2_\alpha\, \hat{Q}^2_\alpha,\ \     \nonumber \\ 
               &=& \left(\hat{N}_\alpha +\frac{1}{2}\right)\, \omega_\alpha,\ \ 
 \end{eqnarray} 
a canonical ensemble of Bosonic oscillators in thermal equilibrium at temperature $T_\alpha$ is described by    
\begin{equation}
\label{th}
\hat\rho_{T_\alpha}=\frac{e^{-\hat{H_\alpha}/k\, T_\alpha}}{Z_{\alpha}}=\frac{e^{-\hbar\omega_\alpha/2k\, T_\alpha}}{{Z_{\alpha}}}\,\sum_{n_\alpha=0}^{\infty}\, e^{-n_\alpha\hbar\omega_\alpha/k\, T_\alpha}\, \vert\, n_\alpha\rangle\langle \, n_\alpha\rangle
\end{equation} 
where  
\begin{eqnarray}
\label{partf}
Z_{\alpha}&=&{\rm Tr}[e^{-\hat{H}/kT_\alpha}]= \,{\rm Tr}\left[e^{- \frac{\hbar\omega_\alpha}{\,kT_\alpha}\,\left(\hat{N}_\alpha+\frac{1}{2}\right)}\right] \nonumber \\
			&=&\frac{e^{-\hbar\omega_\alpha/2\,kT_\alpha}}{1-e^{-\hbar\omega_\alpha/kT_\alpha}}
\end{eqnarray} 
denotes the partition function. The single mode thermal state $$\hat\rho_{T_\alpha}~=~\left(1-e^{-\hbar\omega_\alpha/kT_\alpha}\right)\,\displaystyle\sum_{n_\alpha=0}^{\infty}\, 
e^{-n_\alpha\hbar\omega_\alpha/k\, T_\alpha}\, \vert\, n_\alpha\rangle\langle \, n_\alpha \vert$$ may be readily identified with the canonical single mode Gaussian state $\rho_\alpha$ of (\ref{single}) appearing in the Williamson canonical decomposition (\ref{G2mode}).          

The variance matrix ${\bf V}$ of any two-mode thermal state $\hat{\rho}_{AB}=\hat{\rho}_{T_A}\otimes\hat{\rho}_{T_B}$ is given by~\cite{nm1}   

\begin{eqnarray}
\label{vars}
{\bf V}&=& \left(\begin{array}{cc}{\bf V}_A & 0 \\  0 & {\bf V}_B \end{array} \right)  \nonumber    \\
{\bf V}_A&=&\frac{1}{2}\,\coth\left(\frac{\hbar\omega_A}{2k\, T_A}\right)\, \mathbbm{1}_2, \ \ {\bf V}_B=\frac{1}{2}\,\coth\left(\frac{\hbar\omega_B}{2k\, T_B}\right)\, \mathbbm{1}_2 
\end{eqnarray}
where $\mathbbm{1}_2$ denotes $2\times 2$ identity matrix. Note that ${\bf V}$ is in the Williamson normal form, with its symplectic eigenvalues related to the frequency and temperature of the thermal state of oscillator systems $A$ and $B$ as,     
\be 
\label{nuT}
\nu_{T_\alpha}=\frac{1}{2}\, \coth\left(\frac{\hbar\omega_\alpha}{2k\, T_\alpha}\right), \ \alpha=A,B. 
\ee

In the next subsection  we give an outline of some preliminary notions on quantum phase-space formalism  in terms of  Wigner distribution functions.    
  
\subsection{Wigner distribution function of a Gaussian state} 

 Wigner distribution function plays a central role in developing quantum phase-space formalism involving non-commuting canonical observables $\hat{q}_\alpha,\hat{p}_\alpha$.  Wigner representation of a quantum system in phase-space allows one to explore the connection between quantum and classical formalisms. In this subsection we outline some preliminary notions on the Wigner function associated with single mode continuous variable Gaussian quantum system.  

Wigner function  $W(q,p)$ of  a single mode continuous variable quantum state $\hat{\rho}$  is a real function of phase-space canonical variables $q,p$  defined by~\cite{hil}   
\be
\label{wigdefO}
W(q,p)=\frac{1}{\pi \hbar}\,\int_{-\infty}^{\infty}\, dx \langle q-x\vert \hat{\rho} \vert q+x\rangle e^{2\,i\pi px/\hbar} 
\ee 
where  $\vert q \rangle$ denotes eigenvector of the operator $\hat{q}$; it satisfies the normalization property~\cite{hil} 
\be
\label{sum1}
\int_{-\infty}^{\infty} \int_{-\infty}^{\infty} dq\, dp\, W(q,p)  =\mbox{Tr}(\hat{\rho})=1 
\ee 
which is true of any  probability distribution and gives correct marginal probability distributions: 
\begin{eqnarray}
\label{sum2}
\int_{-\infty}^{\infty}dp\,  W(q,p)  =\langle q\vert \hat{\rho}\vert q\rangle, \ \ \ \ 
\int_{-\infty}^{\infty}dq\,  W(q,p)  =\langle p\vert \hat{\rho}\vert p\rangle  
\end{eqnarray}

Quantum expectation value of any operator $\hat{f}(\hat{q},\hat{p})$ in a  state $\hat\rho$ can be replaced by a  phase-space integration using Wigner function  $W(q,p)$
 as,   
\be
\label{expval}
\mbox{Tr}\left[\hat{\rho}\,\hat{f}(\hat{q},\hat{p})\right]=\int_{-\infty}^{\infty} \int_{-\infty}^{\infty}dq\,dp\,  W(q,p)\,f(q,p). 
\ee 
where Weyl's correspondence rule
\be
\label{weyl}
q^k\,p^l\, \longleftrightarrow\, \frac{1}{2^k}\sum_{r=0}^{k}\, \frac{k!}{r!\, (k-r)!}\,\hat{q}^r\, \hat{p}^l\, \hat{q}^{k-r}
\ee
to associate classical functions  with quantum operators has been employed~\cite{hil}. The classical functions $f(q,p)$  are expressed as the {\em Weyl transforms} of the corresponding operators  $\hat{f}(\hat{q},\hat{p})$ as,   
\be
\label{weyl_transform}
f(q,p) = \int_{-\infty}^{\infty}\, dy   \, \langle q-y/2\vert \hat{f}(\hat{q},\hat{p}) \vert q+y/2\rangle\, e^{-i\pi p\, y/\hbar}  
\ee
so that~\cite{hil} 
\be
\label{product_weyl}
\int_{-\infty}^{\infty} \int_{-\infty}^{\infty}dq\,dp\, f(q,p)\, g(q,p) = (2\pi\hbar)\, {\rm Tr}[\hat{f}(\hat{q},\hat{p})\, \hat{g}(\hat{q},\hat{p})]
\ee
holds for the Weyl transforms of the operators $\hat{f}(\hat{q},\hat{p})$ and  $\hat{g}(\hat{q},\hat{p}).$ 
 
While the Wigner-Weyl formalism allows phase-space description (see (\ref{sum1}), (\ref{sum2}), (\ref{expval})) of quantum theory in a classical language,  the {\em weight} function $W(q,p)$ is not necessarily  non-negative  everywhere  and hence it is  termed as     
{\emph{quasi}-probability  distribution function~\cite{hil}.  

Interestingly, the Wigner function associated with single mode quantum Gaussian states given by~\cite{nm1} 
\be
\label{wV}
W(\xi)=\frac{1}{2\pi\,\sqrt{\det{\bf V}}}\, e^{-\frac{1}{2}\,\xi^T\,{\bf V}^{-1}\,\xi} 
\ee
is non-negative everywhere. Here, ${\bf V}$ is the variance matrix of the Gaussian state  and phase-space canonical variables are expressed compactly in the form of a column $\xi=(q,p)^T$.  The Wigner function  of a single mode Gaussian state $\hat\rho_{T}=\frac{e^{-\hat{H}/k\, T}}{Z}$  in thermal equilibrium at temperature $T$ ~\cite{hil} takes the following form (by substituting the variance matrix  of a thermal state (see (\ref{vars})) and after simplification):       
\begin{eqnarray}
\label{wig1}
W(q,p)&=&\frac{1}{2\pi\,\nu_{T}}{\rm 
	exp}\left[-\frac{1}{2\,\nu_{T}}\, (q^2 +p^2) \right]   \\ 
&=& \frac{1}{2\pi\,\nu_{T}}{\rm 
	exp}\left[-\frac{H(Q,P)}{\hbar\omega\, \nu_{T}}\, \right] \nonumber 
\end{eqnarray} 
where $H(Q,P)=\frac{P^2}{2m}+\frac{1}{2}m\omega^2\,Q^2$,\  $Q=\left(\hbar/m\omega\right)^{1/2}\, q,\, P=\left(m\omega\hbar\right)^{1/2}\, p$  and $\nu_{T}~=~\frac{1}{2}\,\coth\left(\frac{\hbar\omega}{2\,k\, T}\right)$ is the symplectic eigenvalue of the variance matrix (see (\ref{nuT})).   
It is evident that $W(q,p)$ of (\ref{wig1}) is a Gaussian function of  $q$, $p$ and is non-negative.

Using the Wigner function description it is possible to develop a time-reversal symmetric phase-space trajectory approach to derive heat exchange flctuation relation analogous to the Jarzynski-W{\'o}zcik approach in the classical scenario (as outlined in Section~2).       

We  discuss heat exchange statistics in two Gaussian systems, in thermal equilibrium at different temperatures in the next section. 

\section{Heat exchange fluctuation theorem for Gaussian thermal states}

Let us consider two quantum systems $A$ and $B$ characterized by their respective Hamiltonians    
\begin{equation}
\label{hahb}
\hat{H}_A(\hat{\xi}_A)=\frac{\hat{P}^{2}_A}{2m_A}+\frac{1}{2}\, m_A\omega_A^2\, \hat{Q}^{2}_{A},\ \ 
\hat{H}_B(\hat{\xi}_B)=\frac{\hat{P}^{2}_B}{2m_B}+\frac{1}{2}\, m_B\omega_B^2\,\hat{Q}^{ 2}_B, 
\end{equation}
where $\hat{\xi}_\alpha=(\hat{q}_\alpha,\hat{p}_\alpha)^T$,   $\hat{q}_\alpha=\sqrt{\frac{m_\alpha\omega_\alpha}{\hbar}}\, \hat{Q}_\alpha$,  $\hat{p}_\alpha=  \frac{1}{\sqrt{m_\alpha\omega_\alpha\hbar}}\, \hat{P}_\alpha,\ \ \alpha=A, B$.

Let the  systems  be prepared in a thermal state  at temperatures $T_A$, $T_B$ respectively i.e.,     
\begin{eqnarray}
\label{t0}
\hat\rho_{T_A}=\frac{e^{-\hat{H}_A(\hat{\xi}_A)/k\, T_A}}{Z_{A}},\ \ 
\hat\rho_{T_B}=\frac{e^{-\hat{H}_B(\hat{\xi}_B)/k\, T_B}}{Z_{B}}. 
\end{eqnarray}
 
The Wigner function  $W(\xi^{\, 0})$ at time $t=0$ corresponding to the two-mode Gaussian thermal state $\hat\rho^{\,0}_{AB}=\hat\rho^{\,0}_{T_A}\otimes \hat\rho^{\,0}_{T_B}$  is a product of Wigner functions $W(\xi^0_A)$,  $W(\xi^0_B)$, where 
$\xi^{\,0}=(\xi_A^{\,0},\xi_B^{\,0})^T$ and  $\xi^{\,0}_A=(q_A^{\,0},p_A^0)^T$, $\xi^{\,0}_B=(q_B^{\,0},p_B^{\,0})^T$ are classical phase-space columns at $t=0$. Using (\ref{wig1}) we obtain   
\begin{eqnarray}
\label{W0}
W({\xi}^{\,0})&=& \frac{1}{(2\pi)^2\, \nu_{T_A}\nu_{T_B}}{\rm 
	exp}\left[-\left(\frac{\xi^{\,0\,T}_A\,\xi^{\,0}_A }{2\nu_{T_A}}\,+\frac{\xi^{\,0\,T}_B\, \xi^{\,0}_B}{2\nu_{T_B}}\right) \right]\, \nonumber \\ 
&=& \frac{1}{(2\pi)^2\, \nu_{T_A}\nu_{T_B}}{\rm 
	exp}\left[-\left(\frac{H_A(\xi^0_A) }{\hbar\omega_A\nu_{T_A}}\,+\frac{H_B(\xi^0_B)}{\hbar\omega_B\nu_{T_B}}\right) \right]
\end{eqnarray}
where $H_\alpha(\xi^0_\alpha)=\frac{P^2_\alpha}{2m}+\frac{1}{2}m\omega^2_\alpha\,Q^2_\alpha$,\  $Q_\alpha=\left(\hbar/m_\alpha\omega_\alpha\right)^{1/2}\, q_\alpha,\, P_\alpha=\left(m_\alpha\omega_\alpha\hbar\right)^{1/2}\, p_\alpha,$\ $\alpha~=~A,B$. 
The systems $A$ and $B$ are kept in contact with each other in terms of  a {\em quadratic} interaction Hamiltonian (representing a  canonical transformation in phase-space)         
$\hat H_{\rm int}(\hat{\xi^t})$,   which is turned  on and off at time $t=0$,   
$t=\tau$ respectively. 

Corresponding to the unitary time evolution operator
\begin{eqnarray}
\label{Utau}
\hat{U}(\hat{\xi}^\tau)&=&\exp{\left[ -\frac{i\, \tau\,}{\hbar} \left(\hat{H}_A(\hat{\xi}^\tau_A)+\hat{H}_B(\hat{\xi}^\tau_B)+\hat{H}_{\rm int}(\hat{\xi}^\tau)\right)\right]},\   \hat{\xi}^\tau=(\hat{q}^\tau_A,\,\hat{p}^\tau_A\, ; \hat{q}^\tau_B,\, \hat{p}^\tau_B)^T   \nonumber  \\   
\end{eqnarray}
there exists a  $4\times 4$ real symplectic matix~\cite{nm, nm1}   ${\bf S}^\tau\in$Sp(4,R) which acts on the  column of canonical operators as   
 $\hat{\xi}^0=(\hat{q}^0_A,\,\hat{p}^0_A;\, \,\hat{q}^0_B,\,\hat{p}^0_B)^T$ resulting in 
 \be 
\hat{\xi}^0\stackrel{{\bf S}^\tau}{\longrightarrow} \hat{\xi}^\tau={\bf S}^\tau \, \hat{\xi}^0. 
\ee
Consequent to the unitary evolution  of the quantum state $\rho^{\, 0}_{AB}\longrightarrow \rho^{\, \tau}_{AB}~=~\hat{U}(\hat{\xi}^\tau)\,\rho^{\, 0}_{AB}\,\hat{U}^\dag(\hat{\xi}^\tau)$,  the Wigner function undergoes the transformation: $W(\xi^0)~\longrightarrow~W(\xi^\tau)$, where $\xi^\tau~=~(q^\tau_A,\,p^\tau_A\, ; q^\tau_B,\, p^\tau_B)^T$. We thus obtain, 
\begin{eqnarray}
\label{Wtau}
W({\xi}^\tau)&=& \frac{1}{(2\pi)^2\, \nu_{T_A}\nu_{T_B}}{\rm 
	exp}\left[-\left(\frac{H_A(\xi^\tau_A)}{\hbar\omega_A\, \nu_{T_A}}\,+\frac{H_B(\xi^\tau_B)}{\hbar\omega_B\, \nu_{T_B}}\right) \right]\, 
\end{eqnarray}

Under time-reversal operation one finds that  $\hat{\xi}\longrightarrow \hat{\xi}^*=(\hat{q}_A,\, -\hat{p}_A\,; \hat{q}_B,\, -\hat{p}_B)^T$. The unitary dynamics  is chosen to be  invariant under time-reversal i.e.,    
\begin{eqnarray}
\label{qtr}
\hat{H}_A(\hat{\xi}_A)\longrightarrow \hat{H}_A(\hat{\xi}^*_A) = \hat{H}_A(\hat{\xi}_A) \nonumber \\
\hat{H}_A(\hat{\xi}_B)\longrightarrow \hat{H}_B(\hat{\xi}^*_B) = \hat{H}_B(\hat{\xi}_B)  \\
\hat{H}_{\rm int}(\hat{\xi})\longrightarrow  \hat{H}_{\rm int}(\hat{\xi}^*)=\hat{H}_{\rm int}(\hat{\xi}). \nonumber
 \end{eqnarray} 
Consider dynamical  evolution of the system under time-reversal,  transforming the phase-space column of observables   $\hat{\xi}^{0\,*}=\hat{\bar{\xi}}^{\,\tau}$ to  $\hat{\xi}^{\tau\,*}=\hat{\bar{\xi}}^{\,0}$.  Ratio of the Wigner functions 
$W(\xi^0)/W(\bar{\xi}^0)$ is then  given by  (see  (\ref{W0}) and (\ref{Wtau})), 

\begin{eqnarray}
\label{ratio}
\frac{W({\xi}^0)}{W(\bar{\xi}^0)}&=&  \frac{W({\xi}^0)}{W(\xi^{\tau\,*})} \nonumber \\ 
&=& \exp\left[\left(\frac{H_A(\xi^{\,\tau\,*}_A)-H_A(\xi^{\,0}_A)}{\hbar\omega_A\, \nu_{T_A}}\right)\right]\,\exp\left[\left(\frac{H_B(\xi^{\,\tau\,*}_B)-H_B(\xi^{\,0}_B)}{\hbar\omega_B\, \nu_{T_B}}\right)\right] \nonumber \\ 
&=& \exp\left[\frac{\bigtriangleup\,E_A}{\hbar\omega_A\, \nu_{T_A}}\right]\, \exp\left[\frac{\bigtriangleup\,E_B}{\hbar\omega_B\, \nu_{T_B}}\right] 
\end{eqnarray}
where  
\begin{eqnarray}
\label{p1}
\bigtriangleup E_A&=& H_A(\xi^{\,\tau\, *}_A)-H_A(\xi^{\,0}_A),\ \   
\bigtriangleup E_B= H_B(\xi^{\,\tau\,*}_B)-H_B(\xi^{\,0}_B). 
\end{eqnarray}  
Considering a  weak interaction $\hat{H}_{\rm int}(\hat{\xi}^\tau)$, it is deduced that
\begin{eqnarray}
\label{en_con}
H_A(\xi^{0}_A) + H_B(\xi^{0}_B)\approx H_A(\xi^{\tau*}_A) + H_B(\xi^{\tau*}_B) \nonumber \\ 
\Rightarrow \bigtriangleup\, E_A \approx - \bigtriangleup\, E_B.  
\end{eqnarray}
implying  that the net change in internal energy of system $A$ is compensated  by an opposite change in the internal energy of system $B$, when the phase-space trajectory $\xi^t$ is sampled. In other words,  heat transfer during forward realization  ${\cal Q}(\xi^0)=\bigtriangleup\, E_B$ is opposite to that of the reverse realization i.e.,  ${\cal Q}(\bar{\xi}^0)=\bigtriangleup\, E_A=-{\cal Q}(\xi^0)$. Thus, we obtain    
\begin{equation}
\label{ratio2}
\frac{W(\xi^0)}{W(\bar{\xi}^0)}=e^{\bigtriangleup\beta_\omega\, {\cal Q}({\xi}^0)}, 
\end{equation}
where 
\begin{eqnarray}
\label{deltabq}
\bigtriangleup\beta_\omega&=& \beta_{B\,\omega}-\beta_{A\,\omega} \nonumber \\ 
&=& \frac{1}{\hbar\omega_B\, \nu_{T_B}}-\frac{1}{\hbar\omega_A\, \nu_{T_A}} \nonumber \\ 
                  &=& \frac{2\, \tanh\left(\frac{\hbar\omega_B}{2\,k\,T_B}\right)}{\hbar\omega_B } - \frac{2\, \tanh\left(\frac{\hbar\omega_A}{2\,k\,T_A}\right)}{\hbar\omega_A.\, }
\end{eqnarray}
Heat distribution $p_\tau({\cal Q})$ can then be expressed in terms of the Wigner function as 
\begin{eqnarray}
\label{final1}
p_\tau({\cal Q})&=&\int {\rm d}\xi^0\,  W(\xi^{\,0})\, \delta({\cal Q}-{\cal Q}(\xi^{\,0}))\nonumber \\
&=& e^{\bigtriangleup\beta_\omega\,{\cal Q} } \int {\rm d}\bar\xi^{\, 0}\, W(\bar{\xi}^{\,0})\, \delta({\cal Q}+{\cal Q}(\bar\xi^{\,0}))\nonumber \\ 
&=&  e^{\bigtriangleup\beta_\omega\,  {\cal Q}}\, p_{\tau}(-{\cal Q}). 
\end{eqnarray}
We obtain
\begin{eqnarray}
\label{jwqo}
\ln\left(\frac{p_\tau({\cal Q})}{p_\tau(-{\cal Q})}\right)=\bigtriangleup\beta_\omega\, \,{\cal Q}  
\end{eqnarray}
where $\bigtriangleup\beta_\omega$ is given by (\ref{deltabq}). We thus arrive at  a  Jarzynski-W{\'o}zcik {\em like} heat exchange fluctuation relation (\ref{jwqo}) for Gaussian thermal states in the quantum scenario.

\section{Discussions}

Some relevant discussions on  different forms (\ref{eq1}), (\ref{jwq}), (\ref{jwqo}) of the Jarzynski-W{\'o}zcik heat exchange fluctuation relations are summarised in the following.    

\begin{itemize}
\item[(i)] In the classical limit $\hbar\rightarrow 0$ we get $\bigtriangleup\beta_\omega\rightarrow \bigtriangleup\beta$.  Thus the  heat exchange fluctuation relation  (\ref{jwqo}) reduces to its classical analogue (\ref{eq1}) in this limit.    We draw attention to Ref.~\cite{r11} where Wigner phase-space formalism  was employed  to derive the characteristic function (which is the Fourier transform of the probability distribution) of  {\em quantum} work. It was shown that the characteristic function for the  {\em classical} work is recovered in the limit $\hbar\rightarrow 0$. Furthermore,  a generalized Jarzynski identity for {\em quantum} work has been derived very recently~\cite{r12} using the Wigner-Weyl  phase-space approach, which retrieves the celebrated {\em classical work} identity~\cite{j97,j97a} in the classical limit $\hbar\rightarrow 0$. It is for the first time that we have derived a Jarzynski-W{\'o}zcik {\em like} heat exchange fluctuation relation (\ref{jwqo}) for a system of two  quantum harmonic oscillators, which reduces to the classical relation (\ref{eq1}) in the limit $\hbar\rightarrow 0$.
 \item[(ii)] Note that   ${\cal Q}(\xi^0)=H_B(\xi^{\,\tau\, *}_B)-H_B(\xi^{\,0}_B)$, is the {\em classical heat}  exchanged along the  phase-space  trajectory, which appears in the Wigner-Weyl  formalism. However, this formal phase-space trajectory has no clear physical interpretation  because of the uncertainty relation in the quantum realm. Description of the quantum evolution in terms of classical trajectory is feasible in the limit  $\hbar\rightarrow 0$. This explains the transition of  heat exchange fluctuation relation (\ref{jwqo}) to its classical counterpart (\ref{eq1}).

It would be interesting to explore the meaning of the heat distribution $p_\tau({\cal Q})$ of (\ref{final1}) in the Wigner-Weyl formalism. To this end, let us denote a quantum operator $\hat{p}_\tau({\cal Q})$ corresponding to the heat distribution $p_\tau({\cal Q})$. The Weyl transform of the  operator $p_\tau({\cal Q})$ is then identified   to be   the  Dirac delta function (see (\ref{final1})) 
$\delta({\cal Q}-{\cal Q}(\xi^{\,0}))$. Following (\ref{weyl_transform}), we  may  express
\begin{eqnarray}
\label{weyl_pQ}
\delta({\cal Q}-H_B(\xi^{\tau *}_B)+H_B(\xi^{0}_B)) &=& \int d\mathbf{y}    \langle {\mathbf{q-y}/2}\vert \hat{p}_\tau({\cal Q}) \vert \mathbf{q+y/2}\rangle e^{-i\pi (\mathbf{p\cdot y}/\hbar},  \nonumber \\ 
\end{eqnarray}
where $  \mathbf{q}\equiv(q_A,\,q_B),\ \mathbf{p}\equiv(p_A,\,p_B)$, and $\mathbf{y}\equiv(y_A,\,y_B).$
The matrix element $\langle \mathbf{q'}\vert \hat{p}_\tau({\cal Q}) \vert \mathbf{q''}\rangle,\ \ \mathbf{q}=(\mathbf{q'+q''})/2=(q_A,q_B)$ of the operator  $\hat{p}_\tau({\cal Q})$ is then given by the inverse Weyl transform 
$$\langle \mathbf{q-y}/2\vert \hat{p}_\tau({\cal Q}) \vert \mathbf{q+y}/2\rangle=\frac{1}{(2\pi\hbar)^2}\, \int\, d\mathbf{p}\,  \delta({\cal Q}-H_B(\xi^{\,\tau\,*}_B)+H_B(\xi^{\,0}_B))\,
 e^{i\pi \mathbf{p\cdot y}/\hbar}.$$
In other words, the matrix element $\langle \mathbf{q'}\vert \hat{p}_\tau({\cal Q}) \vert\mathbf{q''}\rangle$ is the inverse Fourier transform of the delta function $\delta({\cal Q}-H_B(\xi^{\,\tau\,*}_B)+H_B(\xi^{\,0}_B))$. This prompts us to identify that the operator $\hat{p}_\tau({\cal Q})$  behaves like a (plane wave) projector
such that the trajectory
	starting from the phase-space point $\left(\mathbf{q},\mathbf{p}\right)$ corresponds to an exchange of heat ${\cal Q}=H_B(\xi^{\,\tau\,*}_B)-H_B(\xi^{\,0}_B)=H_A(\xi^{\,0}_A)-H_A(\xi^{\,\tau\,*}_A)$ between the systems $A$ and $B$.
  However, as mentioned earlier,  this  trajectory approach is hindered by the underlying uncertainty relation in the quantum scenario, though such a representation can be validated in the classical limit $\hbar\rightarrow 0$. This explains the reduction of  Jarzynski-W{\'o}zcik heat exchange fluctuation relation (\ref{jwqo}) to its classical analogue (\ref{jwq}) in the limit  $\hbar\rightarrow 0$ where a {\em legitimate}  interpretation of the phase-space trajectories is possible. 
\item[(iii)] The moment generating function~\cite{rmp9,pre18} associated with the heat 
distribution $p_\tau({\cal Q})$  in the Wigner-Weyl formalism may be constructed as follows: 
\begin{eqnarray}
\label{mgfn1}
\left\langle\,e^{-s\, \bigtriangleup\beta_\omega\,{\cal Q}}\right\rangle&=&G_\tau(\bigtriangleup\beta_\omega; s)= \int {\rm d}{\cal Q}\, p_\tau({\cal Q})\, e^{-s\, \bigtriangleup\beta_\omega \, {\cal Q}} \nonumber \\
&=& \int {\rm d}\xi^0\,   W(\xi^{\,0})\,\left\{ \int {\rm d}{\cal Q}\, e^{-s\, \bigtriangleup\beta_\omega\, {\cal Q}}  \delta({\cal Q}-{\cal Q}(\xi^{\,0})) \right\}\nonumber \\ 
	&=&  \int {\rm d}\xi^0\,   W(\xi^{\,0})\, e^{-s\, \bigtriangleup\beta_\omega\, {\cal Q}(\xi^{\,0})} \nonumber \\ 
	&=&  \int {\rm d}\xi^0\,   W(\xi^{\,0})\, e^{-s\, \bigtriangleup\beta_\omega\,\left[ H_B(\xi^{\,\tau\,*}_B)-H_B(\xi^{\,0}_B)\right]}.  
\end{eqnarray}
where $\bigtriangleup\beta_\omega$ is defined in (\ref{deltabq}) and $s$ is an arbitrary real parameter. Substituting (\ref{W0}),(\ref{Wtau}),(\ref{p1}), (\ref{en_con}),  and simplifying, we obtain
\begin{eqnarray}
\label{renyi}
\left\langle\,e^{-s\, \bigtriangleup\beta_\omega\,{\cal Q}}\right\rangle
&=& \frac{1}{(2\pi)^2\, \nu_{T_A}\nu_{T_B}} \int {\rm d}\xi^0\,  
	e^{-\left[\beta_{A\omega}\, H_A(\xi^0_A)+ \beta_{B\omega}\,H_B(\xi^0_B)\right]}    \nonumber \\
	&& \hskip 0.5in   \times 	e^{-s\beta_{B\omega}\, \left[\, H_B(\xi^{\,\tau*}_B)-H_B(\xi^0_B)\right]}\,  e^{s\beta_{A\omega}\,\left[ -H_A(\xi^{\,\tau\,*}_A)+H_A(\xi^{\,0}_A)\right]}\nonumber \\ 
	&=& \frac{1}{(2\pi)^2\, \nu_{T_A}\nu_{T_B}} \int {\rm d}\xi^0\, 	\left\{e^{-\left[\beta_{A\omega}\, H_A(\xi^0_A)+ \beta_{B\omega}\,H_B(\xi^0_B)\right]}\right\}^{1-s}\nonumber \\ 
&& \hskip 0.6in \times	\left\{e^{-\left[\beta_{A\omega}\, H_A(\xi^{\,\tau*}_A)+ \beta_{B\omega}\,H_B(\xi^{\,\tau*}_B)\right]}\right\}^s \nonumber \\ 
&=& \int {\rm d}\xi^0\,  \left[ W(\xi^{\,0})\right]^{1-s}\, \left[W(\xi^{\,\tau})\right]^{s}   \nonumber \\   
 &=&  {\rm exp}\left[(1-s)\, R_s\left(W^\tau\vert\vert W^0 \right)\right],
\end{eqnarray}
where 
\be
R_s\left(W^0\vert\vert W^\tau \right)=\frac{1}{1-s}\,\ln\,\left\{\int {\rm d}\xi^0\,  \left[ W(\xi^{\,0})\right]^{1-s}\, \left[W(\xi^{\,\tau})\right]^{s}\right\}
\ee
 denotes the order-$s$ R{\'e}nyi divergence between the Wigner functions $W(\xi^{\,0})$ (corresponding to the initial state) and $W(\xi^{\,\tau})$ (representing the final state) of the total system $A$ and $B$.

Substituting  $s=1$ in (\ref{renyi}),  we obtain   
\begin{eqnarray}
\label{s1}
\left\langle e^{-\, \bigtriangleup\beta_\omega\, {\cal Q}} \right\rangle=1. 
\end{eqnarray}	

Applying Jensen inequality $\left\langle e^{-\, \bigtriangleup\beta_\omega\, {\cal Q}}\right\rangle \geq  e^{-\,\bigtriangleup\beta_\omega\, \langle\, {\cal Q}\rangle}$ in (\ref{s1}),  we get 
\be
\label{sec_law}  
 \bigtriangleup\beta_\omega\,\langle {\cal Q}\rangle \geq  0 \Rightarrow  \left(\beta_{B\omega}-\beta_{A\omega}\right)\, \langle {\cal Q}\rangle \geq 0.
\ee
Let us assume that $\beta_{B\omega}>\beta_{A\omega}$ (which is analogous, in the limit $\hbar\rightarrow 0$, to the condition $T_A > T_B$). 
We thus obtain a variant of the Clausius inequality (the second law of thermodynamics)
\be
\left(\beta_{B\omega}-\beta_{A\omega}\right)\, \langle {\cal Q}\rangle \geq 0
\ee 
which implies that   heat does not flow  from  system $B$ (cold)  to system  $A$ (hot).      

 In the low temperature limit $T_A,T_B\rightarrow 0$, we obtain (see  (\ref{deltabq}), (\ref{jwqo}))
	\be 
	\label{q0}
	\frac{p_\tau({\cal Q})}{p_\tau(-{\cal Q})}={\rm exp}\left[\frac{2{\cal Q}}{\hbar}\, \left(\frac{1}{\omega_B}-\frac{1}{\omega_A}\right)\right]
	\ee
which indicates that there is still a finite heat flow. In other words, (\ref{q0}) points out that the heat-exchange statistics is not symmetric about $T_A=T_B$, when the frequencies of the oscillators $A$ and $B$ are not equal. In this context, it is of interest to note that Chimonidou and Sudarshan~\cite{ecgs} had investigated {\em relaxation phenomena} of a system of two harmonic oscillators in thermal equilibrium, prepared initially  at different temperatures $T_A$, $T_B$. They subjected the system  to a specific (symplectic) interaction Hamiltonian for a fixed time duration $\tau$ repeatedly, till the system approaches equilibrium. They concluded that the equilibrium reached, when the frequencies of the oscillators are unequal, is {\em not} a thermal one. It is of interest to investigate the asymmetry of the Jarzynski-W{\'o}zcik relation (\ref{jwqo}) at $T_A=T_B$ of a system of  two oscillators with unequal frequencies evolving under different quadratic interaction Hamiltonians. A detailed exploration on these aspects will be reported in a separate communication.    

 In  Ref.~\cite{pre18} Wei had established a connection between the moment generating function $G_\tau(\bigtriangleup\beta; s)$ in the quantum regime  with  the order-$s$ R{\'e}nyi divergence $R_s\left(\rho^{\,0}_{AB}\vert\vert\rho^{\,\tau}_{AB}\right)=\frac{1}{1-s}\,\ln\,\left\{ {\rm Tr}[\left(\rho^{\,\tau}_{AB}\right)^{1-s}\, \left(\rho^{\,0}_{AB}\right)^s]\right\}$  between
	the initial, final density operators $\rho^{0}_{AB}$, $\rho^{\tau}_{AB}$: 
\begin{equation}
\label{qRenyi}
 G_\tau(\bigtriangleup\beta; s)
=\int {\rm d}{\cal Q}\, p_\tau({\cal Q})\, e^{-s\, \bigtriangleup\beta \, {\cal Q}}= {\rm exp}\left[(1-s)\, R_s\left(\rho^{\,0}_{AB}\vert\vert\rho^{\,\tau}_{AB}\right)\right]
\end{equation}
 To arrive at the relation (\ref{qRenyi}) the double projection measurement approach  (as proposed by Jarzynski and W{\'o}zcik~\cite{jw}) was employed. It may be seen that there is a close resemblance between the generating functions  (\ref{renyi}) and (\ref{qRenyi}), which are derived using different approaches. In (\ref{qRenyi})  the generating function $G_\tau(\bigtriangleup\beta; s)$ is related to  the order-$s$ R{\'e}nyi divergence $R_s\left(\rho^{\,0}_{AB}\vert\vert\rho^{\,\tau}_{AB}\right)$ between the density operators $\rho^{0}_{AB}$, $\rho^{\tau}_{AB}$,  whereas (\ref{renyi}) derived using the Wigner phase-space formalism -- in the specific example of harmonic oscillator system --  connects the moment generating function $G_\tau(\bigtriangleup\beta_\omega; s)$ with the R{\'e}nyi divergence $R_s\left(W^0\vert\vert W^\tau \right)$ between the Wigner functions $W(\xi^0),\, W(\xi^\tau)$.   Comparing  (\ref{renyi})  and (\ref{qRenyi}) for the system of two harmonic oscillators, by subjecting the system to specfic symplectic interaction Hamiltonians $\hat{H}_{\rm int}(\hat{\xi}^\tau)$ would be useful for exploring the nature of deviation of  the heat flow statistics in the double projective measurement method  and the Wigner-Weyl phase-space approach.

\item[(iv)] Equipartition theorem plays a fundamental role in  classical statistical physics. It states that for a system in thermal equilibrium at temperature $T$ the average energy per degree of freedom is given by $\frac{1}{2}\, k\, T.$  Equipartition theorem of energy holds universally in classical statistical physics as  it neither depends on the number of particles in the ensemble nor on the nature of the potential acting on the particles. For a system of one dimensional classical harmonic oscillators, in thermal equilibrium at temperature $T$, contribution to the average energy comes from mean kinetic energy and mean potential energy i.e.,  $\langle E\rangle=k\, T$. It has been pointed out~\cite{QEqP1, QEqP2,  QEqP3}   recently that the classical energy equipartition theorem does not hold  in the quantum realm. It is seen that the  average energies  in the state $\rho_{AB}=\rho_{T_A}\otimes \rho_{T_B}$ (see  (\ref{t0})) -- characterizing  a system of quantum harmonic oscillators  $A$ and $B$ in thermal equilibrium at temperatures $T_A,\ T_B$ respectively --  are given by     
	\begin{eqnarray}
	\label{haE}
	\langle \hat{H}_{A}\rangle&=& {\rm Tr}[\rho_{T_A}\, \hat{H}_A]\nonumber \\         
	                          &=& \frac{\hbar\omega_A}{2}\, \left(\langle \hat{q}^2_A \rangle + \langle \hat{p}_A^2 \rangle \right)\nonumber \\ 
	                          &=& \frac{\hbar\omega_A}{2}\, \coth\left(\frac{\hbar\omega_A}{2\,k\,T_A}\right)=\hbar\omega_A\nu_{T_A}  \\
\label{hbE}
	\langle \hat{H}_{B}\rangle&=& {\rm Tr}[\rho_{T_B}\, \hat{H}_B]\nonumber \\         
	&=& \frac{\hbar\omega_B}{2}\, \left(\langle \hat{q}^2_B \rangle + \langle \hat{p}_B^2 \rangle\right) \nonumber \\ 
		&=& \frac{\hbar\omega_B}{2}\, \coth\left(\frac{\hbar\omega_B}{2\,k\,T_B}\right) = \hbar\omega_B\nu_{T_B}                           
	\end{eqnarray} 
where we have made use of (\ref{cov}), (\ref{vars}) and (\ref{hahb}). 

Here the average energies (\ref{haE}), (\ref{hbE}) depend on frequencies $\omega_A,\omega_B$ (indicative of the nature of the potential)     besides  temperatures $T_A,\, T_B$.  The factor $\bigtriangleup\,\beta_\omega$ in the Jarzynski-W{\'o}zcik heat exchange fluctuation relation (\ref{jwqo}) approaches its classical analogue  $\bigtriangleup\,\beta$ of (\ref{eq1}) only in the  limit $\hbar\rightarrow 0$ (see (i) above).  Deviations of  (\ref{jwqo}) from (\ref{eq1}) could be attributed  to the fact that  classical energy equipartition  is no longer valid in the quantum scenario. A series of recent papers~\cite{QEqP1,QEqP2,QEqP3} have proposed quantum counterpart of energy equipartition theorem, which may shed more light on the quantum heat exchange statistics of thermal harmonic oscillator system.    

\item[(v)] It is pertinent to point out some recent results on Jarzynski-W{\'o}zcik XFT, where quantum-to-classical transition is studied.  Denzler and  Lutz~\cite{Lutz18} arrived at  the heat distribution associated with the infinite dimensional systems of thermal
quantum harmonic oscillators, weakly coupled to a heat reservoir  
at  different temperatures, by exactly solving the quantum master equation. The heat distribution so  obtained leads to  the Jarzynski-W{\'o}zcik XFT (\ref{jwq}) for the ensemble of quantum harmonic oscillators in equilibrium. Authors of  Ref.~\cite{Lutz18} implement double measurements on the discrete energy levels of the quantum  thermal oscillator system  at initial time $t=0$ and final time  $t=\tau$ to derive the  XFT (\ref{jwq}) in the quantum regime -  as prescribed originally by  Jarzynski-W{\'o}zcik~\cite{jw}. It is shown that the discrete heat distribution  becomes continuous and  reduces to the corresponding classical expression in the  limit $\hbar\omega/kT\rightarrow 0$. This  falls in line  with the Wigner-Weyl phase-space description on the quantum to classical  transition. 

Highlighting that the quantum features get destroyed in the  double projective measurement scheme (prescribed in the derivation of  quantum Jarzynski-W{\'o}zcik relation (\ref{jwq})), an  entirely different strategy based on dynamic Baysean networks has been employed very recently~\cite{prl20} to derive a {\em fully}  quantum fluctuation theorem for heat exchange in a correlated bipartite thermal system. In the absence of quantum  correlations,  the classical Jarzynski-W{\'o}zcik XFT is recovered from this  quantum fluctuation relation and it is also  shown to reduce to the fluctuation relation in the presence of classical correlations derived in Ref.~\cite{pre15}. 

 It is of interest to note that in Refs.~\cite{r21,r22} direct measurement on the quantum system was avoided by coupling the system to  a classical apparatus. With the  aid of the apparatus the probability distribution for the {\em work done} on a quantum system was constructed in Ref.~\cite{r22} and it was shown that the associated statistics is consistent with the  work fluctuation relation~\cite{j97,j97a,c99}. It is of interest to carry out a detailed study  comparing the heat-exchange fluctuation relation (\ref{jwqo}) for the system of two harmonic oscillators, derived using the Wigner phase-space formalism,  with the ones realized based on different strategies~\cite{prl20, r21, r22}, where direct measurement on the quantum system was judiciously evaded.             
	
\end{itemize}   

In summary, we believe that the Wigner-Weyl phase-space framework to explore Jarzynski-W{\'o}czik XFT 
opens up new perspectives towards understanding the heat distribution statistics and
its clasical limit.

\section*{Acknowledgements}  Sudha and ARU acknowledge financial support from the Department of Science and Technology(DST), India through Project No. DST/ICPS/QUST/Theme-2/2019 (Proposal Ref. No. 107).  AMJ  thanks DST, India for  J.C.Bose National fellowship.  We thank the Referees for their insightful suggestions and for pointing out references \cite{r11,r12,r21,r22}.



\vskip 2cm 

\section*{References}
\begin{thebibliography}{0}
\bibitem{j97} Jarzynski C 1997 Nonequilibrium Equality for Free Energy Differences {\it Phys. Rev. Lett} {\bf 78} 2690--2693 
\bibitem{j97a} Jarzynski C 1997 Equilibrium free-energy differences from nonequilibrium measurements: A master-equation approach 
{\it Phys. Rev.} E {\bf 56} 5018--5035 
\bibitem{c98} Crooks G E 1998  Nonequilibrium Measurements of Free Energy Differences for Microscopically Reversible Markovian Systems, 
{\it J.Stat.Phy.} {\bf 90}  1481--1487 
\bibitem{c99} Crooks G E 1999 Entropy production fluctuation theorem and the nonequilibrium work relation for free energy differences  {\it Phys. Rev.} E {\bf 60} 2721--2726  
\bibitem{jw} Jarzynski C and  W{\'o}zcik D K 2004 Classical and Quantum Exchange Theorems for heat exchange {\it Phys. Rev. Lett.} {\bf 92} 230602
\bibitem{phqe} Jarzynski C 2007 Comparison of far-from-equilibrium work relations {\it C. R. Phys} {\bf 8} 495--506 
\bibitem{Dhar07} Saito K and Dhar A 2007 Fluctuation Theorem in Quantum Heat Conduction, {\it Phys. Rev. Lett.} {\bf 99}, 180601
\bibitem{xft09} Andrieux D,  Gaspard P,  Monnai T and Tasaki S 2009 The fluctuation theorem for currents in open quantum systems {\it New. J. Phys} {\bf 11} 043014 
\bibitem{rmp9} Esposito M, Harbola U and  Mukamel S 2009 Nonequilibrium fluctuations, fluctuation theorems and counting statistics in quantum systems {\it Rev. Mod. Phys} {\bf 81} 1665-1702 
\bibitem{ph10} Campisi M, Talkner P and H{\"a}nggi P  2010 Fluctuation Theorems for Continuously Monitored Quantum Fluxes {\it Phys. Rev. Lett} {\bf 105}, 140601 
\bibitem{rmp11} Campisi M, H{\"a}nggi P and  Talkner P, 2011 Quantum fluctuation relations: Foundations and applications  {\it Rev. Mod. Phys} {\bf 83} 771--791 
\bibitem{pre12} Cohen D and Imry Y 2012 Straightforward quantum-mechanical derivation of the Crooks fluctuation theorem and the Jarzynski equality {\it Phys. Rev.} E {\bf 86} 011111 
\bibitem{prjp12} Rana S, Lahiri S, Jayannavar A M 2012 Quantum Jarzynski equality with multiple measurement and feedback for isolated systems 
{\it Pramana J. Phys} {\bf 79} 233--241 
\bibitem{prx13} Deffner S and Jarzynski C 2013 Information processing and the second law of thermodynamics {\it Phys. Rev.} X {\bf 3} 041003
\bibitem{prjp13} Rana S, Lahiri S, Jayannavar A M 2013 Generalized entropy production fluctuation theorems for quantum systems 
{\it Pramana J. Phys} {\bf 80} 207--222  
\bibitem{epjb14} Lahiri S, Jayannavar A M 2014 Exchange fluctuation theorems for a chain of interacting particles in presence of two heat baths 
{\it Eur. Phys. J.} B {\bf 87} 141 
\bibitem{ijp15} Lahiri S, Jayannavar A M 2015 Derivation of not-so-common fluctuation theorems {\it Indian J. Phys} {\bf 89} 515--523  
\bibitem{ppnature15}  H{\"a}nggi P and  Talkner P 2015 The other QFT {\it Nature Physics} {\bf 11} 108--110
\bibitem{pre15} Jevtic S, Rudolph T, Jennings D, Hirano Y, Nakayama S and Murao M 2015 Exchange fluctuation theorems for correlated quantum systems {\it Phys. Rev.} E {\bf 92} 042113 
\bibitem{pre17}  Pal P S, Lahiri S and Jayannavar A M 2017 Transient Exchange Fluctuation Theorem for heat using a hamiltonian framework {\it Phys. Rev.} E. {\bf 95} 042124 
\bibitem{Lutz18} Denzler T, Lutz E, 2018  Heat distribution of quantum thermal oscillator {\it Phys. Rev.} E {\bf 98} 052106
\bibitem{fqrx18} Alberg J  2018 Fully quantum fluctuation theorems {\it Phys. Rev.} X {\bf 8}, 011019 
\bibitem{pre18}  Wei B B 2018  Relations between heat exchange and R{\'e}nyi divergences  {\it Phys. Rev.} E {\bf 97}, 042107 
\bibitem{rmp20} H{\"a}nggi P and  Talkner P 2019 Statistical mechanics and thermodynamics at strong coupling: Quantum and classical  arXiv:1911.11660v3 
(Reviews of Modern Physics, In press) 
\bibitem{prl20} Micadei K, Landi G T and Lutz E 2020 Quantum fluctuation theorems beyond two-point measurements {\it Phys. Rev. Lett} {\bf 124} 090602 
\bibitem{nm}  Simon R, Mukunda N,  Dutta B 1994 Quantum-noise matrix for multimode systems: U(n) invariance, squeezing, and normal forms {\it Physical Review} A {\bf 49} 1567--1583
\bibitem{nm1} Arvind, Dutta B,  Mukunda N,  Simon R  1995  The real symplectic groups in quantum mechanics and optics  {\it Pramana - J. Phys.} {\bf 45} 471--477
\bibitem{adesso}  Adesso G, Ragy S, Lee R A 2014 Continuous variable quantum information: Gaussian states and beyond, {\it Open. Syst. Inf. Dyn.} {\bf 21} 1440001 
\bibitem{hil} Hillery M,  O'Connell R F, Scully M O,  Wigner E P 1984 Distribution functions in Physics: Fundamentals {\it Phys. Rep.} {\bf 106} 121--167
\bibitem{wil}  Williamson J 1936 On the Algebraic Problem Concerning the Normal Forms of Linear Dynamical Systems {\it Am. J. Math.} {\bf 58} 141--163
\bibitem{r11}   Qian Y,  Liu F 2019 Computing characteristic functions of quantum work in phase space {\it Phys. Rev.} E {\bf 100} 062119 (1--10)
\bibitem{r12}   Brodier O, Mallick K,  Ozorio de Almeida A M 2020 Semi-classical work and quantum work identities in Weyl representation {\it J. Phys. A: Math. Theor.  {\bf 53} 325001}
\bibitem{ecgs}  Chimonidou A, Sudarshan E C G 2008  Relaxation phenomena in a system of two harmonic oscillators {\it  Phys. Rev.}  A {\bf 77} 032121 (1--11)
\bibitem{QEqP1} Bialas P, Spiechowicz J, Luczka J 2018 Partition of energy for a dissipative quantum oscillator {\it Sci. Rep}  {\bf 8} 16080(1--12)
\bibitem{QEqP2} Bialas P, Spiechowicz J, Luczka J 2019 Quantum analogue of energy equipartition theorem  {\it J. Phys.} A {\bf 52} 15LT01
\bibitem{QEqP3} Luczka J 2020 Quantum counterpart of classical equipartition of energy {\it J. Stat. Phys.} {\bf 179} 839–845 
\bibitem{r21}  Yu. V. Nazarov, Kindermann M 2003 Full counting statistics of a general quantum mechanical variable {\it Eur. Phys. J.} B {\bf 35}, 413-420
\bibitem{r22}  Utsumi Y, Golubev D S,  Marthaler  M,  Sch{\" o}n G,   Kobayashi K 2012 Work fluctuation theorem for a classical circuit coupled to a quantum conductor {\it Phys. Rev.} B {\bf 86}, 075420 (1--8)   
\end{thebibliography}
\end{document}